\def\year{2022}\relax
\documentclass[letterpaper]{article} % DO NOT CHANGE THIS
\usepackage{aaai22}  % DO NOT CHANGE THIS
\usepackage{times}  % DO NOT CHANGE THIS
\usepackage{helvet}  % DO NOT CHANGE THIS
\usepackage{courier}  % DO NOT CHANGE THIS
\usepackage[hyphens]{url}  % DO NOT CHANGE THIS
\usepackage{graphicx} % DO NOT CHANGE THIS
\usepackage{natbib}  % DO NOT CHANGE THIS AND DO NOT ADD ANY OPTIONS TO IT
\usepackage{caption} % DO NOT CHANGE THIS AND DO NOT ADD ANY OPTIONS TO IT
\DeclareCaptionStyle{ruled}{labelfont=normalfont,labelsep=colon,strut=off} % DO NOT CHANGE THIS
\usepackage{algorithm}
\usepackage{algorithmic}

\usepackage{newfloat}
\usepackage{listings}
\floatstyle{ruled}
\newfloat{listing}{tb}{lst}{}
\floatname{listing}{Listing}

\usepackage{soul}
\usepackage{xspace}
\usepackage[utf8]{inputenc}
\usepackage{amsmath}
\usepackage{amsthm}
\usepackage{amssymb}
\usepackage{bm}
\usepackage{booktabs} % \midrule
\usepackage{tabularx}% \usepackage{algorithm}
\usepackage{subcaption} % subfigure
\usepackage{multirow}
\usepackage{array}

\newcolumntype{C}[1]{>{\centering\let\newline\\\arraybackslash}m{#1}}

\usepackage{tikz}
\usepackage{stackengine}
\def\checkmark{\tikz\fill[scale=0.4](0,.35) -- (.25,0) -- (1,.7) -- (.25,.15) -- cycle;}

\newcommand{\Eq}{Eq.\@\xspace}

\newcommand{\Fig}{Fig.\@\xspace}

\newcommand{\Tab}{Table\@\xspace}

\newcommand{\Alg}{Algorithm\@\xspace}

\newcommand{\todo}[1]{\textcolor{red}{#1}}
\newcommand{\prelim}[1]{\textcolor{blue}{#1}}

\renewcommand{\todo}[1]{{#1}}
\renewcommand{\prelim}[1]{{#1}}

\title{Label-efficient Hybrid-supervised Learning for Medical Image Segmentation}

\author {
    Junwen Pan\textsuperscript{\rm{1,2}}\footnote{Junwen Pan and Qi Bi contributes equally to the first authorship. This work is conducted when Junwen Pan serves as a research intern at ByteDance Inc.},
    Qi Bi\textsuperscript{\rm{3}}\footnotemark[1]
    ,
    Yanzhan Yang\textsuperscript{\rm{1}},
    Pengfei Zhu\textsuperscript{\rm{2}},
    Cheng Bian\textsuperscript{\rm{1}}\footnote{Corresponding author: biancheng@bytedance.com.}\\
}
\affiliations {
    \textsuperscript{\rm{1}}Xiaohe Healthcare, ByteDance, Guangzhou, China\\
    \textsuperscript{\rm{2}}College of Intelligence and Computing, Tianjin University, Tianjin, China \\
    \textsuperscript{\rm{3}}School of Remote Sensing and Information Engineering, Wuhan University, Wuhan, China\\
    \{junwenpan, zhupengfei\}@tju.edu.cn,
    q\_bi@whu.edu.cn, 
    \{yangyanzhan.yyz, biancheng\}@bytedance.com
}

\begin{document}

\maketitle

\begin{abstract}
Due to the lack of expertise for medical image annotation, the investigation of label-efficient methodology for medical image segmentation becomes a heated topic.
Recent progresses focus on the efficient utilization of weak annotations together with few strongly-annotated labels so as to achieve comparable segmentation performance in many unprofessional scenarios. 
However, these approaches only concentrate on the supervision inconsistency between strongly- and weakly-annotated instances but ignore the instance inconsistency inside the weakly-annotated instances, which inevitably leads to performance degradation.
To address this problem, we propose a novel label-efficient hybrid-supervised framework, which considers each weakly-annotated instance individually and learns its weight guided by the gradient direction of the strongly-annotated instances, so that the high-quality prior in the strongly-annotated instances is better exploited and the weakly-annotated instances are depicted more precisely.
Specially, our designed dynamic instance indicator (DII) 
realizes the above objectives,
and is adapted to our dynamic co-regularization (DCR) framework further to alleviate the erroneous accumulation from distortions of weak annotations. 
Extensive experiments on two hybrid-supervised medical segmentation datasets demonstrate that with only 10$\%$ strong labels, the proposed framework can leverage the weak labels efficiently and achieve competitive performance against the 100$\%$ strong-label supervised scenario.

\end{abstract}

\section{Introduction} \label{sec:intro}

\begin{figure}[!t]
    \centering
	\includegraphics[width=1.0\linewidth]{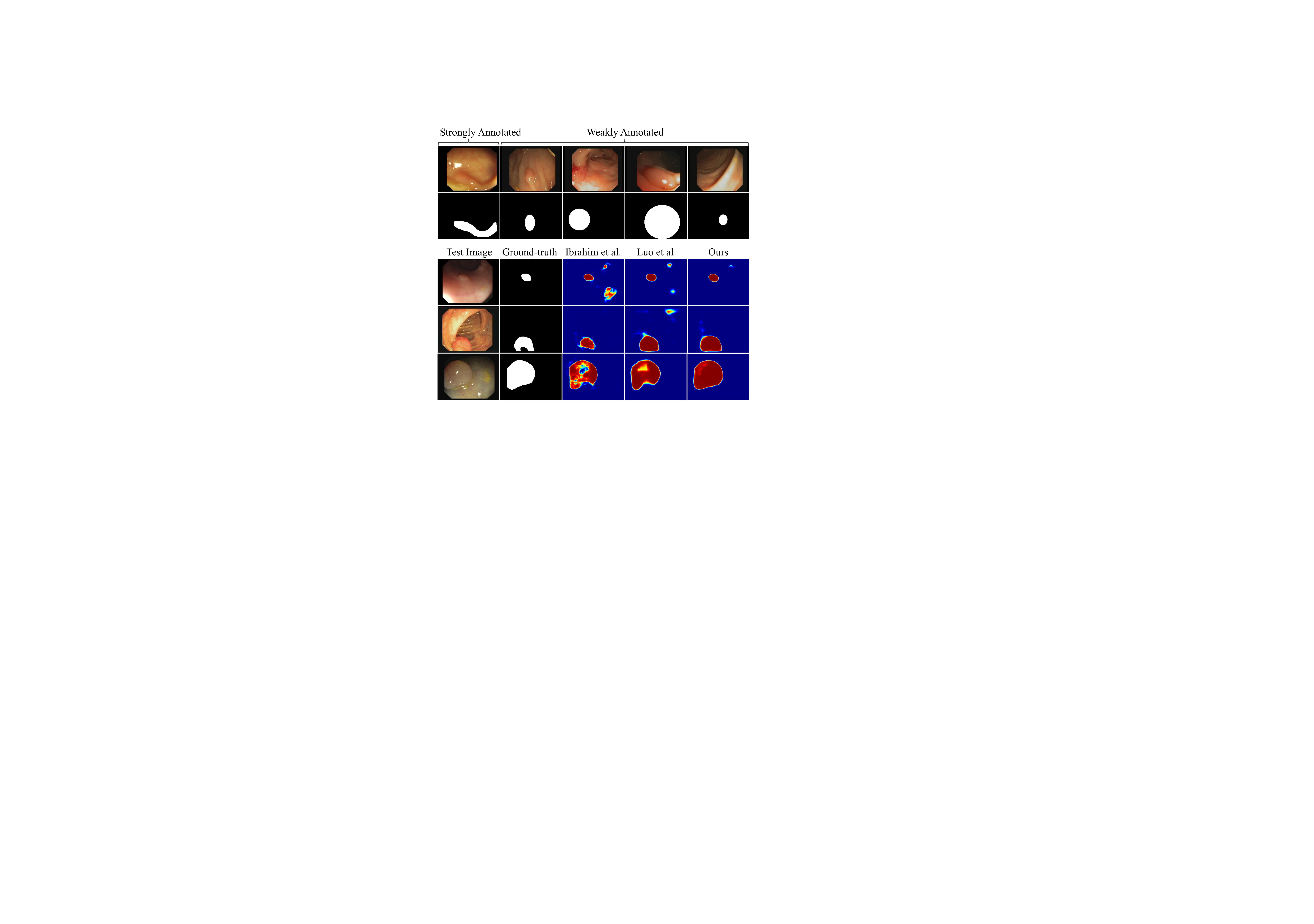}
    \caption{
    \textbf{Top:} Strongly and weakly-annotated instances of endoscopic images and polyp annotations.
    \textit{There are varied distortion levels in the weak annotations, e.g., fairly good (col. 2), slightly distorted (col. 3-4), and completely incorrect (col. 5).}
    \textbf{Bottom:} Test instances and probability maps produced by self-correcting network~\cite{IbrahimVRM20:SelfCorrect}, strong-weak network~\cite{LuoY20:StrongWeak} and our method. %See \Fig\ref{fig:qualitative} for more visual comparisons.
    }
	\label{fig:teaser}
\end{figure}
Medical image segmentation has always been a fundamental task in various biomedical applications, aiming at identifying critical anatomic or pathological structures for further statistical analysis. 
Although significant improvements have been made in recent works~\cite{olaf2015:Unet}, the performance of the deep learning models is strongly impacted by the extensive high-quality annotations, which are expertise-demanding, labor-intensive, and time-consuming, thus hindering deep learning technology from the real-world clinical usages.

Recent studies~\cite{BearmanRFL16:pointsup,LinDJHS16:ScribbleSup} on weakly supervised semantic segmentation demonstrate that weak annotations (e.g., image-level labels, bounding boxes, scribbles) also have the competence of extracting the generalized features compared with strong annotations. Thus, a flurry of techniques~\cite{WeiXSJFH18:mdc,LeeKLLY19:FickleNet} have been explored to utilize only a few pixel-wise strong labels as guidance to learn fine-grained representations for segmentation.
However, as all the strongly- and weakly-annotated instances are treated equally in such solutions, a large number of pseudo labels with less precise depiction capability inevitably dominate the training process, leading to an inferior and less reliable performance~\cite{LuoY20:StrongWeak}. %(see \Fig\ref{fig:teaser} (bottom)).

The key issue responsible for the poor performance of these solutions is the inherent \textit{inconsistency} between weak and strong annotations.
Specifically, the weakly-annotated instances contain a large number of low-quality semantic cues, while the strongly-annotated instances provide scarce yet fine-grained priors.
To tackle this issue and utilize both annotations more efficiently, some efforts proposed multi-branch networks~\cite{LuoY20:StrongWeak} or dual networks~\cite{IbrahimVRM20:SelfCorrect,NingBLZYYGWMZ20:MarcoMicro}, which handle two types of supervision separately.
Specifically, they designed specific approach to interact across strongly and weakly supervised branches, such as self-correction module~\cite{IbrahimVRM20:SelfCorrect}, shared backbone~\cite{LuoY20:StrongWeak}, and exponential moving average~\cite{NingBLZYYGWMZ20:MarcoMicro}, which can alleviate the problem of supervision inconsistency to some extent.

However, the mutual information is still hard to compromise across two parallel branches learned from weakly- and strongly-annotated instances, and thus the over-guiding problem raised by a single type of supervision often exists. 
On the other hand, as these multi-branch frameworks holistically tackle the \textit{supervision inconsistency}, the \textit{instance inconsistency} within the weakly-annotated subset is completely ignored.
In fact, weakly-annotated instances usually contain varying degrees of imaging and annotation distortion (see \Fig\ref{fig:teaser}), which can sometimes mislead the representation learning process and thus low down the model's generalization capability.
For example, a completely incorrect annotation from weakly-annotated subset undoubtedly hurts the overall performance, while a slightly distorted one can provide fairly valuable guidance to the model learning. 
In this case, concentrating on supervision inconsistency while ignoring the instance inconsistency makes the overall trade-off between strongly and weakly supervised branches more intractable, leading to undesirable results (see \Fig\ref{fig:teaser} (bottom)).

To address above issues, we propose a novel label-efficient hybrid-supervised framework regarding the medical semantic segmentation task, %, in which each instance will not only be considered separately, but also be trained in the same step so that the potential of different level annotations can be substantially utilized.
which learns a series of dynamic instance indicators (DII) to reweight each weakly-annotated instance individually under the guidance from strongly-annotated instances.
The learned DII estimates the instance importance, \textit{e.g.}, instances with slight distortions should be granted more attention, while instances with severe annotation distortions will be down-weighted.
In this way, rich and valuable semantic cues contained in massive weakly-annotated instances can be efficiently exploited, while the negative impact is mitigated.
The main contributions of this study can be summarized as follows:
\begin{itemize}
\item[1)] To the best of our knowledge, we are the first to reveal the inconsistency among weakly-annotated instances that obstructs the exploitation of weak annotations in a hybrid-supervised semantic setting.% regarding the medical image segmentation task.
\item[2)] To tackle the obstacle of instance inconsistency, we introduce the DII, which learns a separate weight for each of weakly-annotated instances guided by the gradient direction of strongly-annotated instances.
\item[3)] To further exploiting the fruitful semantic clues from noisy weakly-annotated instances, we design a dynamic co-regularized (DCR) architecture with the aid of DII.
DCR provides a powerful regularization effect and consequently helps avoid erroneous accumulation from the distortion within weak annotations.
\item[4)] Extensive experiments and analysis on the CVC-EndoSceneStill and AS-OCT datasets demonstrate that when only a few strongly-annotated instances are given, our framework has the competence to learn from extensive weakly-annotated instances and achieves the performance close to that of the corresponding fully supervised version.
\end{itemize}

\section{Related Work}

\paragraph{Weakly Supervised Semantic Segmentation.}
Weakly supervised semantic segmentation aims to reduce the annotation efforts by leverage low-cost labels~\cite{LinDJHS16:ScribbleSup,dai2015:boxsup,BearmanRFL16:pointsup}.
These approaches are generally optimized the network and the graphical model alternatively and customized to the specific dataset. For instance, region growing~\cite{HuangWWLW18:DSRG}, CRF~\cite{KrahenbuhlK11:CRF} and GrabCut~\cite{RotherKB04:GrabCut} were employed to constrain the segmentation to coincide with object boundaries.
However, since the graphical model requires definite boundaries between objects, the efficacy of such schemes are doubtful when applying to the medical image dataset.

\paragraph{Hybrid-supervised Semantic Segmentation.}
Hybrid-supervised semantic segmentation introduces few strong annotations combined with weak ones, aiming at provide fine-grained guidance and thus improving the segmentation performance~\cite{PapandreouCMY15:SemiWeakEM}.
For example, generative adversarial network based approaches~\cite{SoulySS17:SemiGANSeg} fused pixel-level labeled data and image-level labeled data through adversarial objectives.
Multi-stage approaches bundled the strong supervision and the proxy supervision estimated from weakly supervised models to learn a single network in the last stage~\cite{WeiXSJFH18:mdc,LeeKLLY19:FickleNet}.
However, the equal treatment of inconsistent annotated data may allow weakly-annotated instances overwhelm the strongly-annotated ones limited in sample numbers, which in turn produces worse results~\cite{LuoY20:StrongWeak}.

To separately use weak annotations and strong ones, a self-correcting network was proposed~\cite{IbrahimVRM20:SelfCorrect}. It trained a primary and an ancillary network and fused features from these two networks via a self-correcting module.
More recently
, a strong-weak network ~\cite{LuoY20:StrongWeak} was proposed to use a shared backbone to exploit the joint information.
Marco-micro framework~\cite{NingBLZYYGWMZ20:MarcoMicro} leveraged the uncertainty-aware consistency and the mean-teacher to provide reliable guidance for both marco and micro branches.
Indeed, these subtle dual-branch approaches accompanied by dedicated mutual interactions that avoided overwhelming the handful yet vital minority and consequently led to compelling performance to some extent.
However, as discussed in the previous section, these interactions only address holistic inconsistency between datasets while ignoring instance inconsistency.

\begin{figure*}[!t]
    \def\svgwidth{\linewidth}
    \includegraphics[width=1.0\linewidth]{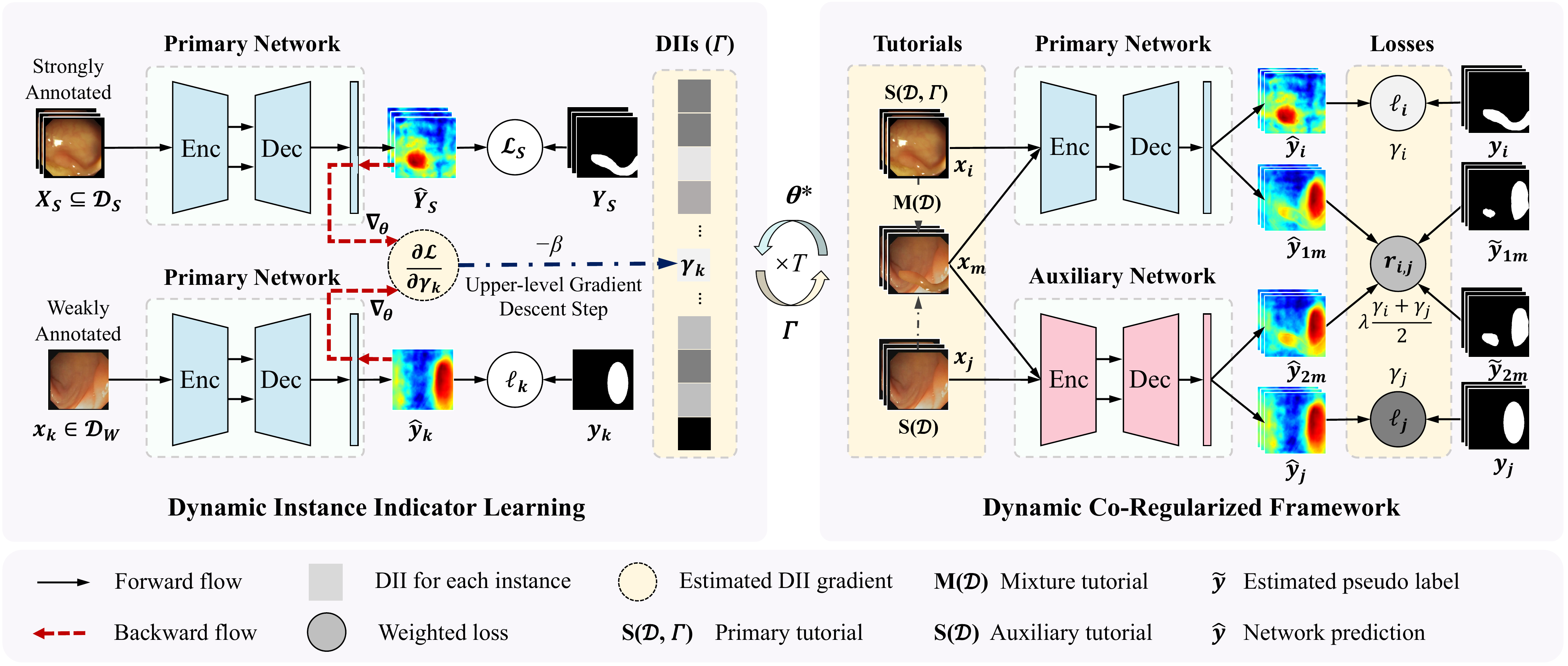}
    % \vspace{-1em}
    \caption{
        Overview of our proposed label-efficient hybrid-supervised learning framework, which alternatively updates the upper-level DIIs ($\Gamma$) and the lower-level DCR networks (${\theta}$).
        \textbf{Left:} Upper-level computation graph, which estimates the upper-level gradient $\frac{\partial \mathcal{L}(\mathcal{D}_S,\theta^*(\Gamma) ) }{ \partial \gamma_k }$ w.r.t. DII $\gamma_k$ guided by the strongly-annotated instances ${X}_S$ (refer to \Eq\ref{eq:approx})  and updates $\gamma_k$ using gradient descent.
        \textbf{Right:} Lower-level dual-network DCR framework, where each network is supervised by a designed tutorial and estimates a pseudo segmentation map to supervise the other network.
    }
    \label{fig:framework}
\end{figure*}

\section{Method}

\Fig\ref{fig:framework} illustrates our proposed hybrid-supervised segmentation framework, which iteratively alternates between upper-level DII and lower-level DCR learning steps.
The upper-level DII learns an individual weight for each weakly-annotated instance via the gradient-based method.
The lower-level DCR framework, on the other hand, employs the dual network structure and is trained in a co-regularized manner, where each network is not only supervised by the corresponding annotation but also by the pseudo mask generated from the other branch.

\subsection{Problem Formulation}

In our hybrid-supervised medical image segmentation, the training set $\mathcal{D}$ consists of a strongly-annotated subset $\mathcal{D}_S=\{ (\mathbf{x}_k, \mathbf{y}_k)| 1\leq k \leq N \}$ with manual pixel-wise delineations and a weakly-annotated subset $\mathcal{D}_W =\{ (\mathbf{x}_k, \mathbf{y}_k)| 1\leq k \leq M \}$ including coarse and noisy masks, where $\mathbf{x} \in \mathbb{R}^{H \times W \times 3}$ is an input image with size $H\times W \times 3$, $\mathbf{y} \in \{0,1\} ^{H\times W \times C}$ is a corresponding mask with $C$ categories, and $N \ll M$.
The core concept of our method is to build a segmentation framework parameterized by $\theta$ and achieves decent performance when learning from massive weakly-annotated instances and minimum strongly-annotated instances. 
We denote the overall objective function of a set of instances $X$ as $\mathcal{L}(X,\theta)$ and the loss of the $i$-th instance as $\ell(\mathbf{x}_i, \mathbf{y}_i, \theta)$. 

\subsection{Dynamic Instance Indicator Learning}
As mentioned before, the distortion of weak annotations introduces the supervision inconsistency and instance inconsistency, posing a major challenge to the efficient and reasonable utilization of weakly- and strongly-annotated data.
To this end, our dynamic instance indicator (DII) is proposed to consider each weakly-annotated instance individually instead of making a global trade-off between weak and strong supervision information like existing methods~\cite{LuoY20:StrongWeak,NingBLZYYGWMZ20:MarcoMicro}.
Intuitively, DII indicates the importance of each weakly-annotated instance and reflects the degree of distortion in a weak annotation.

Specifically, DII comprises a series of learnable indicators $\Gamma=\{\gamma_k | \gamma_k \in [0,1], k \in [1,M] \}$ tailored to $M$ weakly-annotated instances. Note that here we assign a constant value $\mathbf{1}_N$ for $N$ strongly-annotated instances.

To learn such large-scale hyperparameters, we formulate this problem as a bi-level optimization~\cite{dempe2020bilevel} objective, presented as:
\begin{equation}
    \label{eq:bilevel}
    \min_{\Gamma} \mathcal{L}(\mathcal{D}_S,\theta^*(\Gamma)), \; s.t. \; 
    \theta^*(\Gamma) 
    = \arg\min_{\theta} \mathcal{L}(\mathcal{D}, \theta, \Gamma) ,
\end{equation}
where $\mathcal{D}=\mathcal{D}_W \cup \mathcal{D}_S$ denotes the entire dataset, and thus the lower-level loss could be expanded as $\mathcal{L}(\mathcal{D}, \theta, \Gamma)=\frac{1}{N} \sum_{i=1}^{N} \ell(\mathbf{x}_i, \mathbf{y}_i, \theta) + \frac{1}{M}\sum_{k=1}^{M} \gamma_{k}\ell(\mathbf{x}_k, \mathbf{y}_k, \theta)$.
Intuitively, the lower-level objective optimizes the network parameters $\theta$ to minimize the weighted loss over the entire dataset,
while the upper-level objective tries to find the optimal indicators $\Gamma$ based on the performance of strongly-annotated subset $\mathcal{D}_S$.

To solve the above bi-level objectives, we utilize the adaptive gradient descent method~\cite{KingmaB14:Adam} to tune both upper-level DIIs $\Gamma$ and lower-level parameters $\theta$.
Gradient descent methods have been widely adopted for deep network and hyperparameter learning~\cite{Bengio00:GradientBasedOpt,KingmaB14:Adam} and have shown promising performance.
Here, as summarized in \Alg\ref{alg:optimize}, our method iteratively alternates between lower-level network training and upper-level DII updating steps.
At every training iteration, the lower-level step trains the network parameters $\theta$ by gradient descent while fixing the indicators $\Gamma$ (line 4).
After having trained the lower-level network $\tau$ steps, we perform a gradient descent step on the upper-level DIIs $\Gamma$ (line 15).
These two steps are iterated $T$ times and the loss $\mathcal{L}(\mathcal{D}_S,\theta^*(\Gamma) )$ on strongly-annotated subset is expected to reach convergence by adjusting DIIs of weakly-annotated instances.

\paragraph{DII Gradient Estimation.}
The crux of the above algorithm is how to compute upper-level gradient $\frac{\partial \mathcal{L}(\mathcal{D}_S,\theta^*(\Gamma) ) }{ \partial \gamma_k }$ for DII $\gamma_k$ to apply gradient descent updating.
We first decompose the upper-level gradient as:
\begin{equation}
\label{eq:decomposition}
    \frac{\partial \mathcal{L}(\mathcal{D}_S,\theta^*(\Gamma) ) }{ \partial \gamma_k } = \nabla_\theta \mathcal{L} \left(\mathcal{D}_S,\theta^*(\Gamma) \right)^\top \cdot \ \frac{\partial \theta^*(\Gamma)}{\partial \gamma_k} .
\end{equation}
Since $\theta$ is a function resulting from the previous optimization with dependencies on $\Gamma$, the calculation of upper-level gradient w.r.t. $\gamma_k$ requires differentiating the lower-level optimization procedure (\textit{i.e.}, $\arg\min_{\theta} \mathcal{L}(\mathcal{D}, \theta, \Gamma)$), which turns to be intractable in practice.

To tackle the above problem, we then estimate the gradient with the established approximation method~\cite{KohL17:blackbox}.
Assume that $\mathcal{L}$ is second-order differentiable and strictly convex w.r.t. $\theta$, then \Eq\ref{eq:decomposition} can be estimated as: 
\begin{equation}
    \label{eq:approx}
    \frac{\partial \mathcal{L}(\mathcal{D}_S,\theta^*(\Gamma) ) }{ \partial \gamma_k } = 
    % \frac{\partial \mathcal{L}}{ \partial \gamma_k } = 
    - \nabla_{\theta} \mathcal{L} \left(\mathcal{D}_S,\theta^* \right)^\top 
    H_{\theta}^{-1}
    \nabla_{\theta} \ell (\mathbf{x}_k, \mathbf{y}_k, \theta^*) ,
\end{equation}
where $H_{\theta} = \nabla^2_{\theta} \mathcal{L}(\mathcal{D}, \theta^*, \Gamma)$ is the Hessian.

However, it is still impractical to compute \Eq\ref{eq:approx} directly for existing segmentation networks as: 1) computing the Hessian involves second-order gradients, and the complexity of its inverse is far worse than the quadratic one; 2) the update of indicators $\Gamma$ requires the per-instance gradient of the loss function $\mathcal{L}$ w.r.t. huge network parameters $\theta$.

To this end, we utilize several policies to alleviate the computational burden. 
\prelim{First of all, we use identity matrix $I$ to approximate the Hessian, as it is positive definite by assumption~\cite{cook1982residuals}.}
Then, we choose a subset of the network parameters (e.g., parameters from the final layer~\cite{RenYS20:notallunlabeled}) $\theta^\prime \subset \theta$ for gradient computation.
According to the linearity of gradients, the gradient graph of the batch data can be unrolled~\cite{RenZYU18:L2R}, and thus batch acceleration can also be employed.

On the basis of the above approximation, the estimated upper-level gradient $\frac{\partial \mathcal{L}(\mathcal{D}_S,\theta^*(\Gamma) ) }{ \partial \gamma_k }$ actually reflects the ``dissimilarity'' of the network gradients between $k$-th weakly-annotated instance and the strongly-annotated instances.
To be more specific, if the gradient direction from $k$-th weakly-annotated instance is consistent with the strongly-annotated instances, then this instance is beneficial for semantic representation learning and $\gamma_k$ will be larger after performing the gradient descent step.

\begin{algorithm}[t]
    \caption{DII Learning}
    \label{alg:optimize} 
    \begin{algorithmic}[1]
    \REQUIRE strongly-annotated dataset $\mathcal{D}_S$, weakly-annotated dataset $\mathcal{D}_W$, DIIs $\Gamma=\{\gamma_k | \gamma_k \in [0,1], k \in [1,M] \}$, network parameters $\theta$, DII update interval $\tau$, iteration steps $T$, and learning rates $\alpha, \beta$.
    \FOR{$t \leftarrow 1...T$}
    \STATE $X_{batch} \leftarrow \text{BatchSample}(\mathcal{D}_S \cup \mathcal{D}_W) $
    \STATE \textit{// Lower-level (DCR) gradient descent step }
    \STATE $\theta \leftarrow \theta - \alpha \cdot \nabla_{\theta} \mathcal{L}(X_{batch},\theta,\Gamma) $
    \IF{ $(t\mod\tau) \neq 0$}
    \STATE continue
    \ENDIF
    \STATE $\theta^* \leftarrow \theta$ %$, \; \theta^\prime \subset \theta$
    % \STATE \textit{// (2) Update indicators}
    \STATE $X_S \leftarrow \text{BatchSample}(\mathcal{D}_S)$
    \STATE \textit{// Estimate mean gradients on $\mathcal{D}_S$}
    \STATE $\mathbf{g}_S \leftarrow \nabla_{\theta} \mathcal{L}(X_S,\theta^*)$
    \STATE \textit{// Calculate per-instance gradients on $\mathcal{D}_W$}
    \STATE $\mathbf{g}_k \leftarrow \nabla_{\theta} \ell(\mathbf{x}_k, \mathbf{y}_k,\theta^*) ,\;\forall k \in \{1,...,M\}$
    \STATE \textit{// Estimate inverse Hessian matrix}
    \STATE $H^{-1}_{\theta} \leftarrow I$
    \STATE \textit{// Estimate upper-level gradients w.r.t. DIIs}
    \STATE $\frac{\partial \mathcal{L}(X_S, \theta^*{(\Gamma)})}{\partial \gamma_k} \leftarrow -\mathbf{g}_S^\top H^{-1}_{\theta} \cdot \mathbf{g}_k  ,\;\forall k \in \{1,...,M\}$
    \STATE \textit{// Upper-level gradient descent step}
    \STATE $\gamma_k \leftarrow \gamma_k - \beta \cdot \frac{\partial \mathcal{L}(X_S, \theta^*{(\Gamma)})}{\partial \gamma_k} ,\;\forall k \in \{1,...,M\}$
    \ENDFOR
    \end{algorithmic}
\end{algorithm} 

\subsection{Dynamic Co-Regularized Framework}
\todo{
Although DII reweights the instances with different degree of distortions, non-zero weights are assigned to most of weakly-annotated instances.
The distortions in weak annotations can be easily imitated and accumulated with the back-propagation of gradient flows from the deeper to the lower layers~\cite{Araslanov020:SingleStageWSS}.
}
Co-teaching~\cite{HanYYNXHTS18:co-teaching,WeiFC020:JoCoR} and disagreement-based learning~\cite{Yu0YNTS19:howDoesDisagreement} has been utilized to alleviate the noise accumulation from the data label.
Inspired by the similar concept, we propose a dynamic co-regularized (DCR) framework aided by DII to avoid the erroneous accumulation.

\prelim{
As shown in \Fig\ref{fig:framework} (right), DCR is a dual-network architecture, where the primary network has the same architecture as it is in DII. For the auxiliary network, we employ the identical architecture of the primary network but with different initialization.
DCR enforces the consistency on the predictions of two networks, where pseudo labels inferred from one network are used to supervise the other network.
The strength of consistency regularization depends on the disagreement or diversity of two networks~\cite{Yu0YNTS19:howDoesDisagreement}.
To prevent two networks from converging close to each other too quickly, we construct two sampling-based tutorials and a mixture tutorial as our collaborative scheme.
}

\prelim{
In specific, both the primary and auxiliary tutorials contain training pairs drawn from the hybrid dataset $\mathcal{D}$. The primary tutorial samples instances from the multinomial probability distribution corresponding to the DII weights, denoted as $S(\mathcal{D}, \Gamma)$.
In contrast, the auxiliary tutorial uses uniform random sampling.
}
The collaborative training scheme demands an extra input from mixture tutorial (a.k.a, $\mathbf{x}_m$) for the consistency regulation. Here, cutmix~\cite{YunHCOYC19:CutMix} is the optimal option to form the mixture tutorial in our study. We create the mixture tutorial by pasting the foreground regions specified by the given ground truth mask from the image of primary tutorial to the image of auxiliary tutorial. The same operation will be performed consistently to the corresponding pseudo mask as well. More details are described as follows.

Given an image $\mathbf{x}_i$ with annotation $\mathbf{y}_i$ from the primary tutorial, we randomly select a class $c$ in $\mathbf{y}_i$ and extract its binary mask $B^c_i$. The mixed image can be defined as: 
\begin{equation}
    \mathbf{x}_{m} = B^c_i \odot \mathbf{x}_i + (1-B^c_i) \odot \mathbf{x}_j ,
\end{equation}
where $\mathbf{x}_j$ is an image from the auxiliary tutorial and $\odot$ denotes element-wise multiplication.
The image from mixture tutorial $\mathbf{x}_m$ is fed into two networks.
The consistency regulation in DCR constrains two network outputs through the regularization loss:
\begin{equation}
    r_{i,j}=  \ell( \mathbf{x}_{m}, \tilde{\mathbf{y}}_{2m} , \theta_1  ) + \ell( \mathbf{x}_{m}, \tilde{\mathbf{y}}_{1m} , \theta_2 ),
\end{equation}
where $\theta_1$ and $\theta_2$ denote parameters of two networks, and $\tilde{\mathbf{y}}_{1m}$ and $\tilde{\mathbf{y}}_{2m}$ are the pseudo masks generated from two networks. Taking $\tilde{\mathbf{y}}_{1m}$ as an example, we feed  $\mathbf{x}_i$ and $\mathbf{x}_j$ into the primary network and then mix the predictions:
\begin{equation}
    \tilde{\mathbf{y}}_{1m} =  B^c_i \odot \hat{\mathbf{y}}_{1i} + (1-B^c_i)\odot \hat{\mathbf{y}}_{1j},
\end{equation}
where $\hat{\mathbf{y}}_{1i}$ and $\hat{\mathbf{y}}_{1j}$ are network predictions for images $\mathbf{x}_i$ and $\mathbf{x}_j$, respectively.
This pseudo mask is utilized as the supervision for auxiliary network. Similarly, we can obtain the $\tilde{\mathbf{y}}_{2m}$ following the same step.
Eventually, the lower-level loss can be written as:
\begin{equation}
    \begin{aligned}
        \mathcal{L}(\mathcal{D},\theta,\Gamma) = \sum_{i,j} 
        \gamma_i \ell_{i} 
         + \gamma_j  \ell_{j}
         + \lambda \cdot \frac{\gamma_i + \gamma_j}{2}  r_{i,j},
    \end{aligned}
\end{equation}
where $\ell_i$ and $\ell_j$ are losses for primary and auxiliary networks respectively, and $\lambda $ is a hyperparameter to balance the supervised loss and the regularization loss. All losses are implemented with the vanilla pixel-wise cross-entropy loss.

\section{Experiments}
We conduct extensive experiments to verify the effectiveness of our proposed method on different types of medical segmentation tasks with varied types of weak annotations.

\subsection{Datasets and Evaluation Metrics}
\paragraph{Polyp Segmentation.}
The hybrid-supervised polyp segmentation dataset has been built from two public available colonoscopic polyp datasets.
The CVC-EndoSceneStill~\cite{vazquez2017:CVCEndoScene} includes 912 images with elaborately annotated pixel-level labels.
We take 10\% sample pairs ($55$ images) from its training set ($547$ images) as our strong priors and use its test set ($182$ images) for evaluation.
Meanwhile, more than $11k$ frames across $18$ sequences are acquired from the polyp detection dataset CVC-VideoClinicDB~\cite{angermann2017:CVCVid} to provide ellipse masks as weak annotations (refer to \Fig\ref{fig:teaser}).
Although the given weak annotations are trying to approximate the polyp shapes, the inaccurate approximation undoubtedly causes varying degrees of distortions. 
We also introduce artifact of the weak annotation by replacing $40\%$ polyps foreground with the background class so as to simulate unrecognized target during the human annotation.

\paragraph{AS-OCT Segmentation.} 
The hybrid-supervised AS-OCT segmentation dataset is modified from the training set of the Angle closure Glaucoma Evaluation (AGE) Challenge~\cite{AGEData}, which contains over 3200 AS-OCT images with annotations of the closure classification and the coordinates of scleral spurs. 
Same configurations of previous work~\cite{NingBLZYYGWMZ20:MarcoMicro} are adopted in this dataset, where two versions of annotation are entailed. The strong annotation provides the pixel-wise masks of iris and cornea, while the weak annotation is re-annotated with line strokes inside these tissues by experienced ophthalmologists. Then, we follow the same partition protocol in which $60\%$ of the data is used for training, $20\%$ for validation, and the rest $20\%$ for test. It is worth mentioning that only $1\%$ of the training instances utilize strong annotations.

\paragraph{Evaluation Metrics.}

Dice coefficient and average symmetric surface distance (ASSD) ~\cite{HeimannGS09:ASSDEvaluation} are utilized to measure the segmentation performance.
For the polyp segmentation task, the prediction performance is reported on pathological regions.
For the AS-OCT segmentation task, we report above quantitative evaluation metrics regarding the iris and cornea prediction.

\subsection{Implementation Details.}

We implement our algorithm based on the PyTorch framework~\cite{paszke2017:pytorch}.
The DeepLabv3+ structure~\cite{Chieh2018:deeplabv3plus} with a ResNet50 backbone pre-trained on ImageNet~\cite{DengDSLL009:imagenet} is chosen as the primary branch and auxiliary branch in the proposed framework.
Random initialization is applied to the parameters of decoders in two branches.
In addition, $\gamma_1,...,\gamma_M$ is initialized with $0.5$ and clipped to the range of $[0,1]$.
We adopt vanilla Adam optimizer~\cite{KingmaB14:Adam} to tune DIIs with default betas set to 0.9 and 0.999 respectively.
Network parameters are updated iteratively via mini-batch SGD with momentum=0.9, batch size=16 and weight decay=0.00005.
The upper-level and lower-level learning rates are initially set to $0.1$ and $0.002$ by default, respectively.

\begin{table}[!t]
    \centering
    \small
    \begin{tabularx}{1.0\linewidth}{@{}l  X @{} X @{} l c r@{}}
       \toprule
       Model & {DII} & {DCR} & {DII Sampling} & {Dice} & {ASSD} \\
       \midrule
       Baseline & & & & {69.77} &  11.29 \\
    %   \specialrule{\cmidrulewidth}{0pt}{0pt}
    \midrule
       \multirow{4}{*}{Our Variants}
       & \checkmark &  & & {72.52} & 10.46 \\
       & & \checkmark &  & {76.32} & 9.68 \\
       & \checkmark & \checkmark &  & {81.15} & \textbf{8.14} \\
        &\checkmark & \checkmark & \checkmark & \textbf{82.56} & 8.37\\
       \bottomrule
    \end{tabularx}
    \caption{Ablation studies on the polyp segmentation dataset.}
 \label{tab:ablation_main}
 \end{table}

\begin{table}[!t]
    \small
    \centering
    \begin{tabularx}{\linewidth}{@{}X r r@{}}
        \toprule
        DCR Variants & Dice & ASSD \\
        \midrule
        Foreground Paste $\rightarrow$ CutMix & 81.37 & 8.58 \\
        Pseudo Regularization $\rightarrow$  KL Div. & 76.83 & 8.47 \\
        Full DCR & \textbf{82.56} & \textbf{8.37} \\
        \bottomrule
    \end{tabularx}
    \caption{Impact of foreground paste and pseudo regularization strategy on DCR performance.}
    \label{tab:ablation_dcr}
\end{table}

\begin{figure}[t!]
    \centering
    \begin{subfigure}{0.43\linewidth}
        \tiny
       \includegraphics[width=\textwidth]{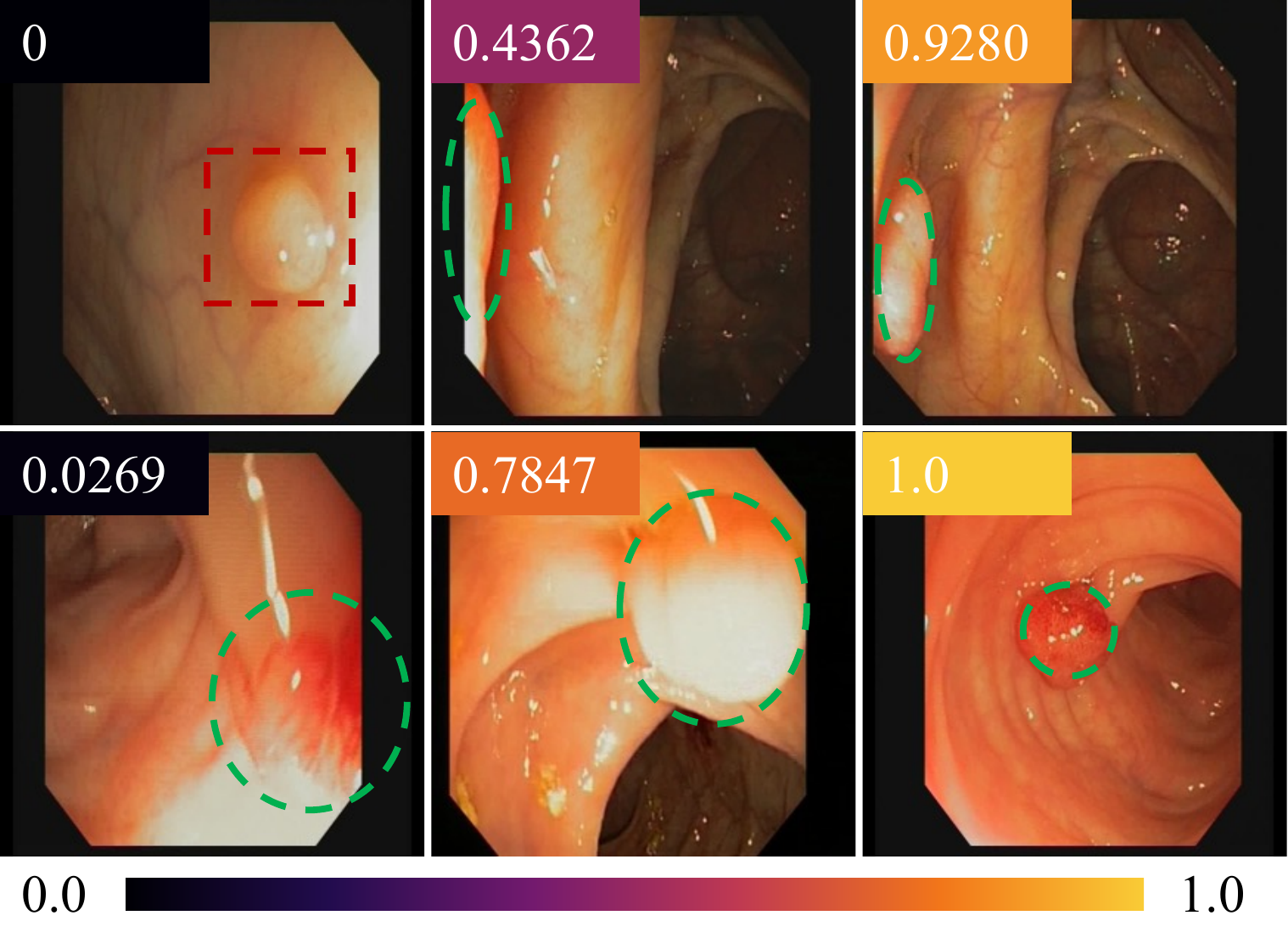}
        \vspace{-1.5em}
       \caption{}
       \label{fig:dii_examples}
    \end{subfigure}
    \begin{subfigure}{0.265\linewidth}
        \tiny
       \includegraphics[width=\textwidth]{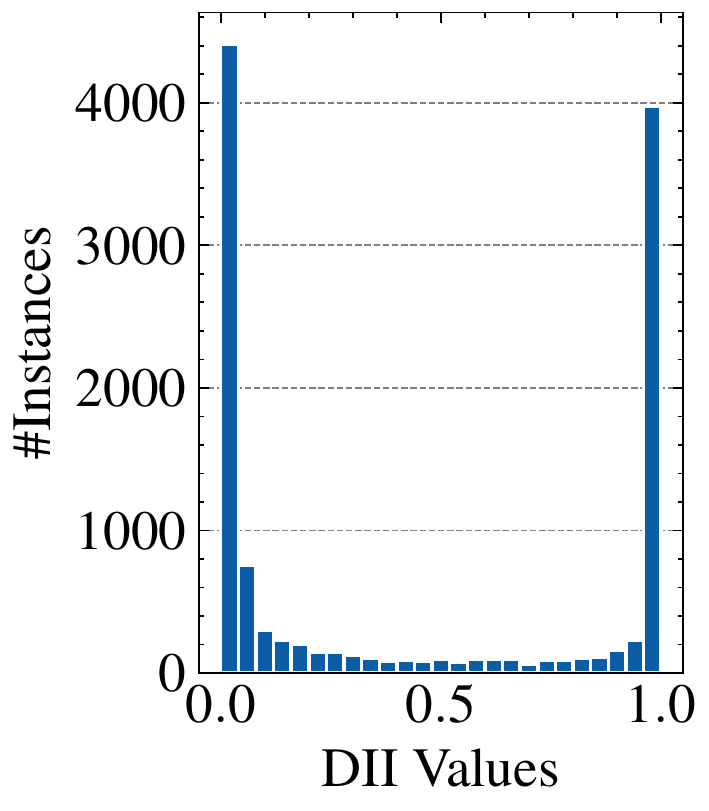}
       \caption{}
       \label{fig:dii_hist}
    \end{subfigure}
    \begin{subfigure}{0.245\linewidth}
        \tiny
        \includegraphics[width=\textwidth]{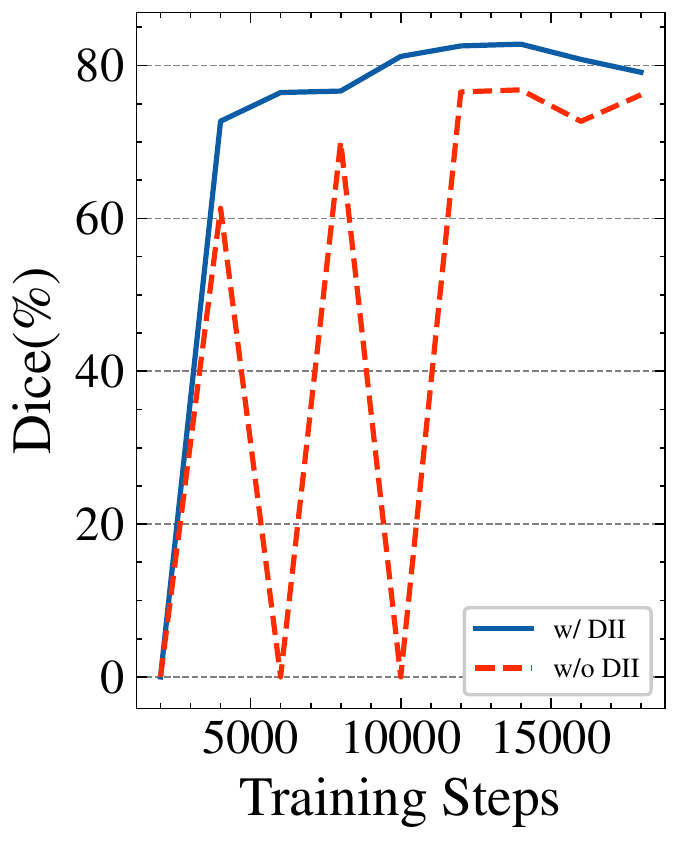}
        \caption{}
        \label{fig:dii_training}
     \end{subfigure}
    \vspace{-0.8em}
    \caption{
        \textbf{(a)}: Demonstration of DIIs for randomly selected instances, where the green ellipses are weak annotations, the red box indicates missing annotation, and the numbers show the optimal DII values. DII successfully reflects  \textit{``importance'' of weakly-annotated instance}. \textbf{(b)}: DII distribution on weakly-annotated polyp segmentation dataset. \textbf{(c)}: Training curves w/ and w/o DII.
    }
\end{figure}

\subsection{Ablation Studies and Analysis}

To verify the effectiveness of the proposed strategies individually, we conduct ablation studies on the polyp segmentation dataset.
Firstly, as performed in Table~\ref{tab:ablation_main}, we roughly merge two types of annotations under the same supervision to train the deep model as our baseline.
Four variants of strategy ablations are designed, including DII, DCR, and DII tutorial. Notable performance gains have been observed when utilizing the proposed DII or DCR individually. 
Furthermore, we achieve 1.41\% extra improvement by applying the DII sampling tutorial additionally, suggesting that our method can effectively utilize both vital strong priors and massive weak semantic cues. 
After that, module ablations are constructed to verify the module effectiveness of DCR. From \Tab~\ref{tab:ablation_main}, we substitute cutmix for foreground paste strategy causing a 1.19\% Dice drop. In addition, using the Kullback-Leibler divergence as the regularization loss results in a 5.73\% performance decrease, demonstrating that the proposed module can effectively exploit noisy semantic cues from weakly-annotated instances.

\paragraph{Understanding DIIs.}

To understand what DII has learned and how it contributes to learning from weak annotations, we conduct experiment on polyp segmentation dataset and investigate its behavior.
We collect learned DII values over all instances from the weakly-annotated set and then visualize several typical instances.
Qualitative results are shown in \Fig\ref{fig:dii_examples}.
DII tries to decrease the importance of instances that contain incorrect annotations and low-quality images, and boost the importance of samples whose weak annotation is close to the ground-truth mask.
We also visualize the histogram in \Fig\ref{fig:dii_hist} to reveal the relationship of the DII values and number of instances.
As can be seen, most instances are pushed toward zero or one value, suggesting that DII tries to make clear useful semantic clues in weak annotations.
Furthermore, \Fig~\ref{fig:dii_training} demonstrates that DII can stabilize the training procedure, since the negative effect raised from noisy annotation has been degraded along with the iterations.

\begin{figure}[t!]
    \centering
    \begin{subfigure}{0.49\linewidth}
        \tiny
       \includegraphics[width=\textwidth]{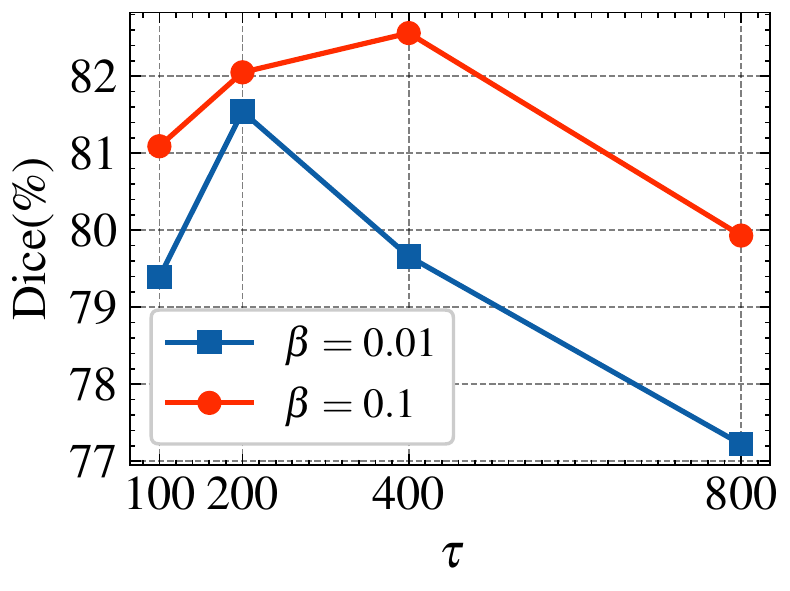}
       \vspace{-2em}
       \caption{Performance with varied combinations of $\beta$ and $\tau$.}
       \label{fig:dii_hyperparameter}
    \end{subfigure}
    \begin{subfigure}{0.49\linewidth}
        \tiny
       \includegraphics[width=\textwidth]{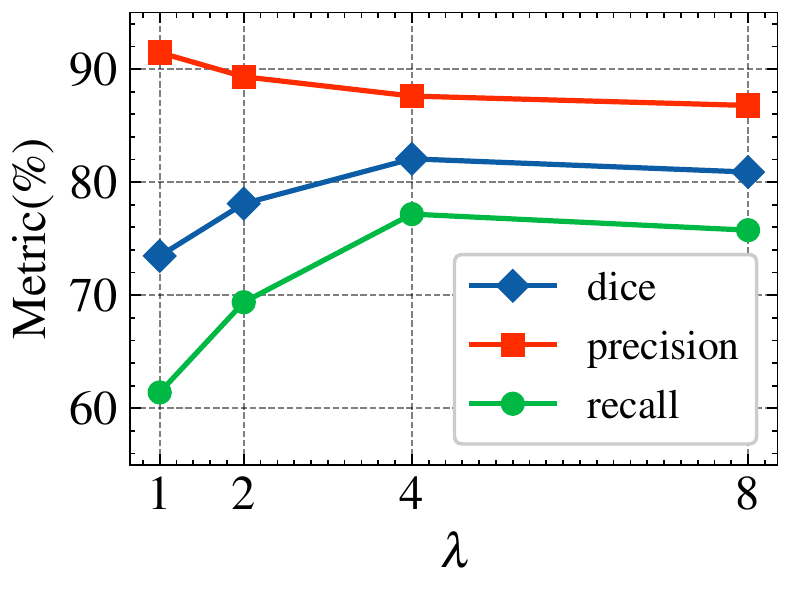}
       \vspace{-2em}
       \caption{Performance with varied regularization strength $\lambda$.}
       \label{fig:lambda}
    \end{subfigure}
    \vspace{-0.3em}
    \caption{ Experiments on different hyperparameters DII (a) and DCR (b) learning.
    }
\end{figure}

\begin{table}[!t]
    \small
    \centering
    \begin{tabularx}{\linewidth}{@{}X c c c c  r@{}}
        \toprule
        \#Strong Annotations  & 28 & 55 & 110 & 220 & 546$^\star$\\
        \midrule
        Dice & {80.36} & {82.56} & {82.97} & {82.99} & {82.91}\\
        \midrule
        ASSD & {8.42} & {8.37} & {7.76} & {7.32} & \text{6.52}\\
        \bottomrule
    \end{tabularx}
    \caption{Experiments on different size of strongly-annotated subset. The asterisk symbol indicates fully supervised baseline (DeepLabv3+) w/o $11k$ weak annotations.}
    \label{tab:num_strong_instances}
\end{table}

\begin{table*}[!t]
    \small
    \centering
    \begin{tabularx}{1.0\linewidth}{l >{\centering\arraybackslash}X@{}>{\centering\arraybackslash}X  >{\centering\arraybackslash}X@{}>{\centering\arraybackslash}X@{}>{\centering\arraybackslash}X@{}>{\centering\arraybackslash}X@{}>{\centering\arraybackslash}X@{}>{\centering\arraybackslash}X@{}}
    \toprule[1pt]
     \multirow{3}{*}{Methods} & \multirow{2}{*}{CVC-EndoSceneStill} & & \multicolumn{6}{c}{AS-OCT} \\
     \cline{4-9}
     &                 &                    & \multicolumn{2}{c}{Cornea} & \multicolumn{2}{c}{Iris} & \multicolumn{2}{c}{Mean}  \\
     \cline{2-9}
       & Dice & ASSD & Dice & ASSD & Dice & ASSD & {Dice} & ASSD  \\
    \hline
    \multicolumn{9}{l}{\textit{Baselines}} \\
    \hline
    Weakly Supervised & 67.43 & {12.66}  &  55.14 & 9.30 & 35.03 & 13.62 & 45.09 & 11.46 \\
    Few Strong Supervised  & 67.67 &  {12.45} &78.83 &5.79	&68.64 &6.29 &73.73	 &6.04 \\
    Fully Supervised  & 82.91 &   {6.52} & 95.71 & 0.13	& 91.59	& 0.21 &{93.65}		&{0.17} \\
    \hline
    \multicolumn{9}{l}{\textit{Hybrid-supervised methods}} \\
    \hline
    FickleNet~\citep{LeeKLLY19:FickleNet}  &  69.77 &  11.29  & 83.71 & 2.58 & 80.55 & 3.11 & 82.13 & 2.85 \\
    Self-Correcting~\citep{IbrahimVRM20:SelfCorrect}  & 67.68 & {12.73} & \textbf{94.16} & 1.52 & 90.36 & 1.43 & 92.26 & 1.47 \\
    Marco-Micro~\citep{NingBLZYYGWMZ20:MarcoMicro} & 72.33 & 10.74 & 93.93 & {1.58} & 89.79 & {1.61} & 91.86 &  {1.60} \\
    StrongWeak~\citep{LuoY20:StrongWeak} & {73.53} &  {14.63} & {--} &{--} &{--} &{--} &{--} &{--} \\
    $\dagger$StrongWeak~\citep{LuoY20:StrongWeak} & {76.41} & 11.42 & 92.15 & 1.85 & 81.68 & 2.62 & 86.91 & 2.24 \\
    \hline
    Ours & \textbf{82.56} & \textbf{8.37} & \textbf{94.39} & \textbf{1.35} & \textbf{91.81} & \textbf{1.19} & \textbf{93.10} & \textbf{1.27} \\
    \bottomrule
    \end{tabularx}
    \caption{Quantitative comparisons with the state-of-the-art methods on the hybrid-supervised polyp segmentation dataset and AS-OCT segmentation dataset.
    ``$\dagger$'' indicates our re-implemented version with strong data augmentation for a fairer comparison.
    }
    \label{table:comparison_result}
\end{table*}

\paragraph{Impact of Hyperparameters of DII and DCR.}

The selection of the hyper-parameters in DII and DCR can influence the learning process.
Initially, we investigate the effect of update interval $\tau$ and learning rate $\beta$ in \Alg\ref{alg:optimize} which balance the learning pace of upper-level DII and lower-level network.
The result is plotted in \Fig~\ref{fig:dii_hyperparameter}. 
Unsurprisingly, we can observe that either too small or large $\tau$ will negatively affect the model performance. 
We finally chose the optimal configuration with  $\tau=400$ and $\beta=0.1$ in our whole study. 

Another experiment is conducted to explore the impact of loss coefficient $\lambda$ of DCR. As illustrated in \Fig\ref{fig:lambda}, it is noticed that the Dice score will increase at the beginning and reach the maximum at $\lambda=4$ and then decrease slowly. In addition, we also present the precision and recall curves at the same time. As the increasing of $\lambda$, we can observe the precision encounters a slight drop but the recall achieves a significant gain. We speculate that the manifestation lies in the effectiveness of the DII-guided tutorial and the foreground paste strategy which encourages the framework to predict the potential regions without overfitting. Therefore, it is inevitable to introduce few false positives that affect the precision.

\paragraph{Impact of the Size of Strongly-annotated Subset.}
The strongly-annotated subset plays two pivotal roles in our framework: serving as a guidance in the upper-level DII learning and providing fine-grained samples in the lower-level training. Hence, the size of strongly-annotated subset on the performance needs to be investigated. 
\Tab\ref{tab:num_strong_instances} shows, unsurprisingly, that increasing strongly-annotated subset consistently improves performance.
On the other hand, we find that an exponentially increase of annotation cost only obtain minor improvement in Dice score, indicating that it is worth developing efficient hybrid-supervised methods to balance annotation budget and performance.

\begin{figure}[!t]
    \centering
	\includegraphics[width=1.0\linewidth]{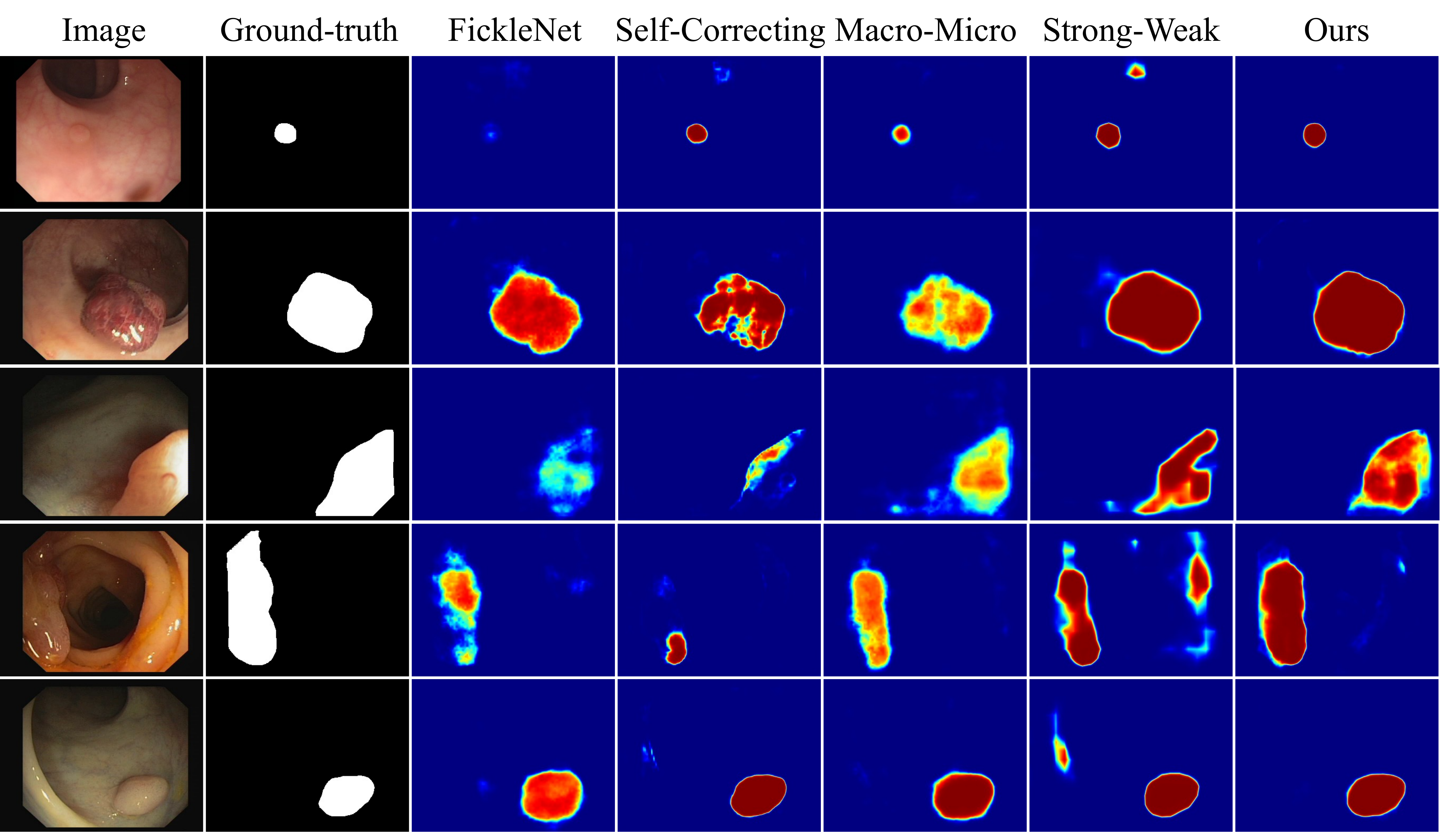}
    \caption{
    Qualitative comparisons of our method with the state-of-the-arts on the polyp segmentation dataset.
    }
	\label{fig:qualitative}
\end{figure}

\subsection{Comparison with State-of-the-arts}
In order to verify the effectiveness of the proposed method, we compare the proposed framework with the state-of-the-art hybrid-supervised semantic segmentation methods.
For fair comparison, all experimental setups are subjected to the identical experimental configurations.
Specifically, three baselines of \textit{Weakly Supervised}, \textit{Few Strong Supervised} and \textit{Fully Supervised} denote a single DeepLabv3+ network~\cite{Chieh2018:deeplabv3plus} trained with only weakly-annotated instances, 10\% strongly-annotated instances, and 100\% strongly supervised instances, respectively. 
\Tab\ref{table:comparison_result} summarizes the experimental results of the proposed method, aforementioned baselines and the compared state-of-the-art methods. 
Our proposed method significantly outperforms all previous approaches by a large margin. 
In particular, our results with only 10\% strongly-annotated training samples are close to that of \textit{Fully Supervised} version, indicating that the fruitful semantic cues in a large number of weakly-annotated instances are exploited sufficiently. 
Moreover, a huge performance discrepancy on two datasets can be observed from all compared methods. 
In contrast, the proposed method consistently yields the best Dice scores among these methods. 
\Fig\ref{fig:qualitative} illustrates several typical qualitative results, where less clutter or incompleteness in our results indicate that our method is more capable of exploiting semantic information in the hybrid annotated dataset.

\section{Conclusion}

This paper proposes a label-efficient hybrid-supervised learning framework for medical image segmentation, which can achieve a competitive performance by exploiting extensive weakly-annotated instances and only a handful of strongly-annotated instances. Specifically, DII learning algorithm and a DCR framework are proposed to extract the useful semantic clues and mitigate the erroneous accumulation during training. DII automatically tunes the weight for each weakly-annotated instance guided by the gradient direction from few strongly-annotated instances, which can assist the framework to overcome the instance inconsistency. 
\prelim{Then, DCR, empowered by the collaborative training scheme and consistency regulation, further relieves the distortion in weak annotations.}
Extensive experiments show that the proposed method substantially outperforms current state-of-the-art approaches and reaches a close performance against the fully supervised scenario. It has great potential to serve as a reliable solution for label-efficient medical image segmentation with limited annotation.

\section{Acknowledgements}
We would like to thank all of the anonymous reviewers for their invaluable suggestions and helpful comments.
This work is supported in part by the National Key Research and Development Program of China under Grant 2019YFB2101904, the National Natural Science Foundation of China under Grants 61732011, 61876127 and 61925602, the Applied Basic Research Program of Qinghai under Grant 2019-ZJ-7017.

\bibliography{aaai22.bib}

\end{document}